\documentclass[aps,pra,nofootinbib,superscriptaddress,reprint]{revtex4-1}
\usepackage[mathlines]{lineno}
\usepackage{blkarray}
\usepackage{mathtools}
\usepackage{bbold}
\usepackage{hyperref}
\usepackage{epsfig}
\usepackage{bm}
\usepackage{amssymb,amsmath,amsfonts,amsthm,dsfont}
\usepackage{latexsym}
\usepackage{graphicx}
\usepackage[caption=false]{subfig}
\usepackage{listings}
\usepackage{float}
\usepackage{enumerate}
\usepackage{array}

\begin{document}
\setlength{\parskip}{0pt}

\title{Noise-tolerant parity learning with one quantum bit}
\author{Daniel K. Park}
\email{kdpspin@gmail.com}
\affiliation{Natural Science Research Institute, Korea Advanced Institute of Science and Technology, Daejeon 34141, Republic of Korea}
\author{June-Koo K. Rhee}
\affiliation{School of Electrical Engineering, Korea Advanced Institute of Science and Technology, Daejeon 34141, Republic of Korea}
\author{Soonchil Lee}
\affiliation{Department of Physics, Korea Advanced Institute of Science and Technology, Daejeon 34141, Republic of Korea}
\begin{abstract}
Demonstrating quantum advantage with less powerful but more realistic devices is of great importance in modern quantum information science. Recently, a significant quantum speedup was achieved in the problem of learning a hidden parity function with noise. However, if all data qubits at the query output are completely depolarized, the algorithm fails. In this work, we present a quantum parity learning algorithm that exhibits quantum advantage as long as one qubit is provided with nonzero polarization in each query. In this scenario, the quantum parity learning naturally becomes deterministic quantum computation with one qubit. Then the hidden parity function can be revealed by performing a set of operations that can be interpreted as measuring nonlocal observables on the auxiliary result qubit having nonzero polarization and each data qubit. We also discuss the source of the quantum advantage in our algorithm from the resource-theoretic point of view.
\end{abstract}
\maketitle
\def\one{{\mathchoice {\rm 1\mskip-4mu l} {\rm 1\mskip-4mu l} {\rm
\mskip-4.5mu l} {\rm 1\mskip-5mu l}}}
\section{Introduction}
\label{sec:intro}
Experimental realizations of quantum information processing (QIP) have made impressive progress in the past years~\cite{Nigg302,NVQEC,SCQEC,NMR2017}. Nonetheless, a scalable architecture capable of universal and reliable quantum computation is still far from within reach. While the development of such quantum computers is pursued, identifying well-defined computational tasks for which less powerful and less challenging devices (for example, subuniversal, without quantum error correction, etc.) can still outperform classical counterparts is of fundamental importance.

One interesting family of problems for which near-term quantum devices can exhibit considerable advantages is machine learning. In particular, the quantum speedup is demonstrated in the problem of learning a hidden parity function defined by the unknown binary string in the presence of noise (LPN). The LPN problem is thought to be computationally intractable classically~\cite{Angluin1988,Blum2003,Lyubashevsky2005,Levieil2006}, and hence cryptographic applications have been suggested based on this problem~\cite{Regev:2005,Pietrzak2012}. In the quantum setting, all possible input binary strings are encoded in the \textit{data} qubits for parallel processing, and the outcome of the function is encoded in the auxiliary \textit{result} qubit. Then the quantum learner with the ability to coherently rotate all qubits before the readout can solve the LPN problem in logarithmic time~\cite{LPNTheory}.
However, the number of required queries diverges as the noise (depolarizing) rate increases, and the learning becomes impossible if the final state of the data qubits is maximally mixed~\cite{LPNTheory,LPNexp}. This result intuitively makes sense since measuring the maximally mixed state outputs completely random bits. Hence the parity function can only be guessed with success probability decreasing exponentially with the size of the problem in both classical and quantum settings.

In this work, we present a protocol with which the hidden bit string of the parity function can be learned efficiently even if all data qubits are completely depolarized, provided that the result qubit has nonzero polarization. Under the aforementioned conditions, the learning algorithm can naturally become \textit{deterministic quantum computation with one quantum bit} (DQC1)~\cite{DQC1PhysRevLett.81.5672}. Then the expectation value measurement on the result qubit allows for efficient evaluation of the normalized trace of the unitary gate that represents the hidden parity function. However, this unitary operator is traceless as long as at least one element of the hidden bit string is 1, and therefore the naive application of the DQC1 protocol does not help. Thus, we modify the original quantum LPN algorithm by adding a set of operations that can be understood as performing nonlocal measurements between each data qubit and the result qubit. With this change, the normalized trace is nonzero if the hidden bit encoded in the data qubit is 0, and zero if it is 1. Therefore, the parity function can be learned using the number of queries that grows only linearly with the length of the hidden bit string. This counterintuitive result shows that the robustness of the quantum parity learning against decoherence is retained via the power of one quantum bit.
The quantum advantage is achieved without any entanglement between the data qubits and the result qubit bipartition. This brings up an interesting question: What is the quantum resource that empowers the learning protocol? We conjecture that the inherent ability of the DQC1 model to discriminate the coherence consumption, which results in producing nonclassical correlation, lies at the center of our learning algorithm.

The remainder of the paper is organized as follows. Section~\ref{sec:2} briefly reviews the LPN problem and the DQC1 protocol, topics that have been studied extensively by numerous authors. In Sec.~\ref{sec:3.1} we describe the equivalence of the quantum parity learning circuit and the DQC1 circuit when the data output is in the maximally mixed state. The DQC1 algorithm for solving the LPN problem is presented in Sec.~\ref{sec:3.2}, including the discussion on the computation efficiency. The effect of errors at various locations in the DQC1 LPN protocol is discussed in Sec.~\ref{sec:3.3}. Section~\ref{sec:4} discusses the origin of the quantum advantage in our learning algorithm, and Sec.~\ref{sec:5} concludes.
\section{Preliminaries}
\label{sec:2}
\subsection{LPN}
\label{sec:2.1}
Here we briefly summarize the work presented in Ref.~\cite{LPNTheory}. In the parity learning problem, an oracle generates a uniformly random input $x\in\lbrace 0,1\rbrace^n$ and computes a Boolean function $f_s$ defined by a hidden bit string $s\in\lbrace 0,1\rbrace^n$,
\begin{equation}
f_s(x)=\sum_{j=1}^n s_jx_j\;\text{mod }2,
\end{equation}
where $s_j\;(x_j)$ is the $j$th bit of $s\;(x)$. A query to the oracle returns $\left(x,\;f_s(x)\right)$, and a learner tries to reconstruct $s$ by repeating the query. In the presence of noise, the learner obtains $ f_s(x)\oplus e$, where $e\in \lbrace 0,1\rbrace$ has the Bernoulli distribution with parameter $p$, i.e., Pr$\lbrack e=1\rbrack=p/2,\;0<p<1$~\cite{Angluin1988,Blum2003,Lyubashevsky2005}. The LPN problem is equivalent to decoding a random binary linear code in the presence of random noise~\cite{Lyubashevsky2005}.

\begin{figure}[h]
\centering
\includegraphics[width=0.9\columnwidth]{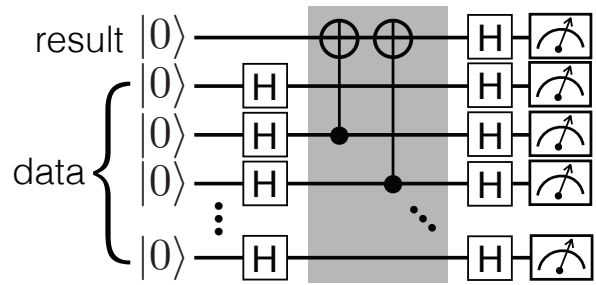}
\caption{\label{fig:LPN0}The quantum circuit for the parity learning algorithm in Ref.~\cite{LPNTheory}. Hadamard operations ($H$) prepare the equal superposition of all possible input states. The hidden parity function (gray box) is realized using controlled-not gates between the result (target) and data (control) qubits. The hidden bit string $s=011\ldots 0$ is used as an example. All qubits are measured in their computational bases at the end.}
\end{figure}

In the quantum setting, the learner has access to a quantum oracle which implements a unitary transformation on the computational basis states and returns the equal superposition of $|f_s(x)\rangle_r|x\rangle_d$ for all possible inputs $x$. The subscript $r\;(d)$ indicates the result (data) qubit. At the query output, the learner applies Hadamard gates to all qubits to create an entangled state
\begin{equation}
\frac{1}{\sqrt{2}}\left(|0\rangle_r|0\rangle^{\otimes n}_d+|1\rangle_r|s\rangle_d\right).
\end{equation}
Thus, whenever the result qubit is $1$ (occurs with probability $1/2$), measuring data qubits in their computational bases reveals $s$. The quantum version of the parity learning is depicted in Fig.~\ref{fig:LPN0}. In this example, $s$ is $011\ldots 0$, and it is encoded via a series of controlled-not (CNOT) gates targeting the result qubit conditioned on the data qubits. The gray box in the figure emphasizes that the learner does not have \textit{a priori} knowledge about this part.
A noisy quantum oracle can be modeled with the depolarizing channel $\mathcal{D}_p\left(\rho\right)=\left(1-p\right)\rho+p\one/2$ acting independently on all qubits at the oracle's output with a constant known noise rate of $p<1$.

Learning from the noiseless oracle is tractable for both quantum and classical learners. However, the quantum algorithm prevails when the noise is introduced. The best-known classical algorithms for LPN  has superpolynomial complexity in $n$~\cite{Angluin1988,Blum2003,Lyubashevsky2005,Levieil2006}, while the quantum learning based on the bitwise majority vote requires $O\left(\log n\right)$ queries and $O\left(n\right)$ running time (gates and measurements)~\cite{LPNTheory}.
This result contradicts the widely accepted idea that quantum computers are intrinsically more vulnerable to error than classical computers. The quantum LPN was realized experimentally with superconducting qubits in Ref.~\cite{LPNexp}.

On the other hand, in terms of the noise strength, the query complexity of the quantum algorithm is $O(\text{poly}(1/(1-p)))$~\cite{LPNTheory}. The experimental results in Ref.~\cite{LPNexp} also show that the number of queries diverges as $p\rightarrow 1$ for both classical and quantum learners. This is evident since maximally mixed states at the query output does not provide any knowledge about $s$. In fact, in the learning algorithm discussed thus far, for each completely depolarized data qubit, the probability of finding $s$ exactly is reduced by $1/2$. Repeating the query does not improve the success probability since the outcome is uniformly random every time. Equivalently, the fully depolarizing noise acting on the data output can be translated to the state preparation error. In this case, without any additional noise, the final entangled state at the end of a query is
\begin{equation}
\frac{1}{\sqrt{2}}\left(|0\rangle_r|x\rangle_d\pm |1\rangle_r|x\oplus s\rangle_d\right),
\end{equation}
where the relative phase factor $-1$ is due to potential error propagations to the result qubit via CNOT gates as the measurement error is moved backward through the quantum circuit. Then the measurement only reveals either $x$ or $x\oplus s$. However, since $x$ is randomly sampled in each query, the learning is not possible. Classically, the complete depolarization of the data qubits at the output corresponds to complete ignorance of the input bit string $x$. Therefore, under such a noise model, classical and quantum learners can only guess $s$ out of $2^{n}$ possibilities.

\subsection{DQC1}
\label{sec:2.2}
DQC1 is a subuniversal quantum computation model to which one probe qubit with nonzero polarization $\alpha$, $n$ bits in a maximally mixed state, an arbitrary unitary transformation, and the expectation measurement of the Pauli operator $\sigma_i$ ($i\in\lbrace x,y,z\rbrace$) on the probe qubit are available~\cite{DQC1PhysRevLett.81.5672}. Though weaker than standard universal quantum computers, it still offers efficient solutions to some problems that are classically intractable~\cite{DQC1PhysRevLett.81.5672,PhysRevA.72.042316,DQC1complexity}. In particular, DQC1 can be employed to efficiently estimate the normalized trace of an $n$-qubit unitary operator, $U_n$, provided that $U_n$ can be implemented with $O(\text{poly}(n))$ elementary quantum gates. In the trace evaluation protocol, a Hadamard gate on the probe qubit is followed by the controlled unitary $\left(|0\rangle\langle 0|\otimes\one_n+|1\rangle\langle 1|\otimes U_n\right)$, where $\one_n$ is the $2^n\times 2^n$ identity matrix. These operations transform the input state $\rho_i=(\one+\alpha\sigma_z)\otimes\one_n/2^{n+1}$ to
\begin{equation}
\rho_f=\frac{1}{2^{n+1}}\left(\one_{n+1}+\alpha\left(|0\rangle\langle 1|\otimes U_n^{\dagger}+|1\rangle\langle 0|\otimes U_n\right)\right).
\end{equation}
The traceless part that deviates from the identity is called the deviation density matrix, and only this part returns nonzero expectation values in DQC1. Measuring the expectation of $\sigma_x$ or $\sigma_y$ on the probe qubit ends the protocol, and
\begin{equation}
\langle \sigma_x\rangle=\frac{\alpha}{2^n}\text{Re}\left(\text{tr}\left(U_n\right)\right),\;\langle \sigma_y\rangle=\frac{\alpha}{2^n}\text{Im}\left(\text{tr}\left(U_n\right)\right).
\end{equation}
Repeating the protocol $O(\log\left(1/\delta\right)/(\alpha\epsilon)^2)$ times allows for estimating the expectation values to within $\epsilon$ with the probability of error $\delta$~\cite{estimate}.
\section{Parity learning with fully depolarizing noise}
\subsection{From LPN to DQC1}
\label{sec:3.1}
\begin{figure}[h]
\centering
\includegraphics[width=0.9\columnwidth]{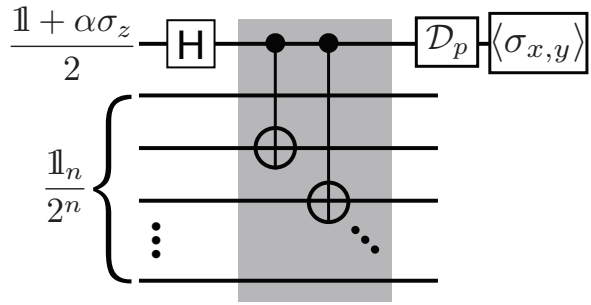}
\caption{\label{fig:LPNasDQC1}The DQC1 circuit that implements the original quantum parity learning algorithm with fully depolarized data qubits. The gray box represents the hidden function defined by $s=011\ldots 0$ as an example. The data qubits at the query input are in the completely mixed state, but the result qubit is prepared with nonzero polarization $\alpha$. The expectation measurement replaces the projective measurement, and it returns the normalized trace of the unitary matrix that corresponds to the hidden parity function. The result qubit also experiences depolarizing noise ($\mathcal{D}_p$) prior to the measurement as in the original algorithm. The trace is zero for all $s$ except when it is uniformly 0.}
\end{figure}

The LPN algorithm fails if the noise completely randomizes the output. However, if the result qubit is alive with some polarization, then can anything about the overall evolution be inferred from measuring the result qubit alone? This situation resembles the DQC1 model in which the probe qubit carries the information about the trace of the unitary operator applied to the completely mixed state.
Indeed, using $H|0\rangle\langle 0|H\otimes\one+H|1\rangle\langle 1|H\otimes \sigma_x=\one\otimes H|0\rangle\langle 0|H+\sigma_x\otimes H |1\rangle\langle 1|H$, the quantum circuit for the parity learning (Fig.~\ref{fig:LPN0}) with completely depolarizing noise on the data qubits can be converted to the DQC1 circuit as depicted in Fig.~\ref{fig:LPNasDQC1}.
Without loss of generality, the data qubits are supplied in the completely mixed state as input. The result qubit can be initialized with some error, but it should possess nonzero polarization. Since measuring the fully depolarized data qubits is redundant, only the result qubit is measured. Then by measuring the expectations of $\sigma_{x}$ and $\sigma_{y}$, the normalized trace of the unitary matrix that implements the hidden parity function can be evaluated. The depolarizing noise at the result qubit ouput scales the expectation values by a factor of $\left(1-p\right)$:
\begin{equation}
\label{eq:trUs}
\langle \sigma_x\rangle+i\langle \sigma_y\rangle=\frac{\left(1-p\right)\alpha}{2^n}\text{tr}\left(U_{s}\right),
\end{equation}
where $U_{s}$ is the unitary implementation of the hidden parity function acting on the data qubits. This is easy to verify using the Kraus representation of the depolarizing channel $\mathcal{D}_p=\lbrace \sqrt{1-3p/4}\one,\sqrt{p/4}\sigma_x,\sqrt{p/4}\sigma_y,\sqrt{p/4}\sigma_z\rbrace$ and the cyclic property of the trace. Hence as long as $p<1$, the normalized trace can be estimated with high accuracy using $\sim 1/(1-p)^2$ repetitions. Equation~\ref{eq:trUs} shows that some information about the hidden function can be contained in the coherent basis of the result qubit. Yet the trace of the hidden unitary matrix does not provide any useful knowledge about $s$ since the trace is zero for all $s$ except when $s$ is uniformly 0.

In the following, we present a strategy for finding $s$ using the trace estimation.
\subsection{Solving LPN using DQC1}
\label{sec:3.2}
\begin{figure}[h]
\centering
\includegraphics[width=0.9\columnwidth]{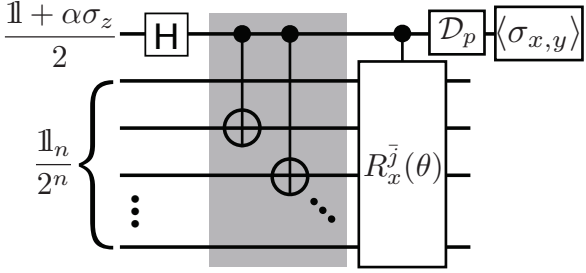}
\caption{\label{fig:modDQC1}The modified DQC1 circuit for solving the LPN problem. The gray box represents the hidden function defined by $s=011\ldots 0$ as an example. The quantum learner applies the uniform rotation to all data qubits controlled by the result qubit except on the $j$th qubit ($R_x^{\bar j}(\theta)$). For some $\theta$, the trace of the total unitary operator is 0 (nonzero) if the hidden bit encoded in the $j$th data qubit is 1 (0).}
\end{figure}

The quantum learner with an access to the DQC1 LPN circuit (Fig.~\ref{fig:LPNasDQC1}) has the ability to implement additional quantum gates after the unknown unitary operation. If a rotation $R_x(\theta)=\exp(i\theta \sigma_x/2)$ controlled by the result qubit is applied equally to all data qubits after the hidden parity function, the trace of the total unitary operator becomes
\begin{equation}
\text{tr}\left(R_x^{\otimes n}(\theta)U_{s}\right)=2^n\left(i\sin\left(\theta/2\right)\right)^m\left(\cos\left(\theta/2\right)\right)^{n-m},
\end{equation}
where $m$ is the number of CNOT operators implemented in the hidden parity function, i.e., the number of ones in $s$.
Now, if the rotation on one of the data qubits is undone by another controlled-rotation $R_x^{j\dagger}(\theta)$,
then the normalized trace of the total unitary operator becomes
\begin{align}
\label{eq:tr_final}
&\text{tr}\left(R_x^{\bar{j}}(\theta)U_{s}\right)/2^n\nonumber\\
&=\begin{dcases}
    \left(i\sin\left(\theta/2\right)\right)^m\left(\cos\left(\theta/2\right)\right)^{n-m-1}&\text{if } s_j=0,\\
    \qquad\quad\qquad\quad\;\; 0&\text{if } s_j=1.
\end{dcases}
\end{align}
Here $R_x^{\bar{j}}(\theta)$ represents a coherent rotation uniformly applied to all $n$ qubits except the $j$th qubit, i.e.,
\begin{align}
\label{eq:cR}
R_x^{\bar{j}}(\theta)&=R_x^{j\dagger}(\theta) R_x^{\otimes n}(\theta)\nonumber \\
&=\exp\left(-i\theta \sigma_x^{(j)}/2\right)\prod_{k=1}^n\exp\left(i\theta \sigma_x^{(k)}/2\right),
\end{align}
and the superscript $(j)$ indicates that the Pauli operator is acting on the $j$th data qubit while the identity operator acts on the rest. We use $\sigma_i^{(r)}$ to represent the result qubit for clarity when needed. For $\theta\neq a\pi,\;a\in\mathbb{Z}$, the DQC1 protocol can resolve whether the hidden bit encoded in the $j$th data qubit is 0 or 1; the trace estimation returns a nonzero value if $s_j=0$, and 0 if $s_j=1$. The quantum circuit for finding the value of $s_j$ is shown in Fig.~\ref{fig:modDQC1}.

The deviation density matrix at the end of the protocol can be written as
\begin{equation}
\tilde\rho_{f}=\frac{\alpha}{2^{n+1}}\left(|0\rangle\langle 1|\otimes V^{\dagger}R_x^{j}(\theta)+|1\rangle\langle 0|\otimes R_x^{j\dagger}(\theta)V\right),
\end{equation}
where $V=R_x^{\otimes n}(\theta)U_s$. Then the expectation measurement on the result qubit can be expressed as
\begin{align}
\label{eq:nonlocal}
\langle \sigma_i\rangle&=\text{tr}\left(\sigma_i^{(r)}\frac{\alpha\cos(\theta/2)}{2^{n+1}}\left(|0\rangle\langle 1|\otimes V^\dagger+|1\rangle\langle 0|\otimes V\right)\right)\nonumber\\
+&\text{tr}\left(\sigma_i^{(r)}\sigma_x^{(j)}\frac{i\alpha\sin(\theta/2)}{2^{n+1}}\left(|0\rangle\langle 1|\otimes V^\dagger-|1\rangle\langle 0|\otimes V\right)\right).
\end{align}
Consequently, the measurement outcome can be interpreted as the sum of two expectations determined from different deviation density matrices, and one of them (second line in Eq.~(\ref{eq:nonlocal})) corresponds to measuring the nonlocal observable on the result qubit and the $j$th data qubit. This nonlocal contribution to the measurement extracts the information about the bit value hidden in the $j$th qubit.

To optimally distinguish the normalized traces (the difference is denoted as $\Delta\tau_j$) without knowing $m$, the rotation angle should be chosen as $\theta=\pi/2$ (or an odd-integer multiple of it). Then $\Delta\tau_j=i^m(1/\sqrt{2})^{n-1}$. Once $s_j$ is revealed, the $j$th data qubit can be decoupled from the result qubit by applying the inverse of the unitary operator that encodes $s_j$. Then in the subsequent run, the controlled rotation is applied only to the remaining data qubits. This rotation can be expressed as
\begin{equation}
\tilde{R}_x^{j}(\theta)=\prod_{k=j+1}^n\exp\left(i\theta \sigma_x^{(k)}/2\right).
\end{equation}
This extra procedure increases $\Delta\tau_j$ by a factor of $\sqrt{2}$ for each decoupled data qubit, i.e., $\lvert\Delta\tau_j\rvert=(1/\sqrt{2})^{n-j}$, and can reduce the computational overhead accordingly.

With these results, the full learning algorithm can be stated as follows.
\begin{itemize}
\item Given a DQC1 circuit with the hidden unitary operator controlled-$U_s$, for $j=1,\ldots ,n$, do the following.
\begin{enumerate}
\item Apply the controlled-rotation $\tilde{R}_x^{j}(\theta)$ to data qubits with the result qubit as the control.
\item Measure $\langle \sigma_x\rangle$ and $\langle \sigma_y\rangle$. Repeat until desired accuracy is reached.
\item If $\langle \sigma_x\rangle+\langle \sigma_y\rangle$=0, record $s_j=1$. Otherwise, record $s_j=0$.
\item If $\langle \sigma_x\rangle+\langle \sigma_y\rangle$=0, apply a bit-flip ($\sigma_x$) gate to the $j$th data qubit controlled by the result qubit. Otherwise, do nothing.
\item Increment $j$ and go to step 1.
\end{enumerate}
\end{itemize}
Until the first nonzero value is detected, both $\langle\sigma_x\rangle$ and $\langle\sigma_y\rangle$ must be measured because $\Delta\tau_j$ can be either real or imaginary depending on $m$. However, once the nonzero trace is found, only one of them needs to be measured in subsequent runs.

The number of queries required for estimating $s$ within $\epsilon$ with the probability of error $\delta$ is $O(\log(1/\delta)/(L(\alpha\epsilon(1-p))^{2}))$, assuming an ensemble of $L$ quantum systems (e.g., spin-1/2 nuclei) encodes the result qubit.
In order to identify whether $s_j$ is 0 or 1 with high certainty, $\epsilon<\lvert\Delta\tau_j\rvert/2$ must be satisfied. On the other hand, $\lvert\Delta\tau_j\rvert^2$ decreases exponentially in $n-j$. Thus, the learning may be too expensive, especially when $L \ll 2^n$ and for $j\ll n$. However, for ensemble quantum computing models such as those based on nuclear magnetic resonance, $L\sim 10^{22}$. This means that for about $n=\log_2(10^{20})\approx 66$, $L\gg 2^n$ and the learning algorithm is efficient.
For the hidden bit string beyond this length, the size of the ensemble should increase exponentially to maintain the efficiency in the number of queries.

\subsection{Error analysis}
\label{sec:3.3}
\begin{figure}[h]
\centering
\includegraphics[width=0.7\columnwidth]{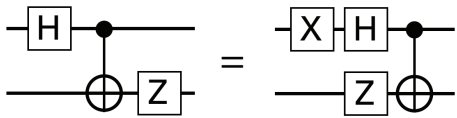}
\caption{\label{fig:ErrorProp} A phase-flip ($Z$) on the second qubit after a gate sequence constructed as a Hadamard followed by a CNOT is equivalent to a bit-flip ($X$) on the first qubit and a phase-flip on the second qubit before the gate sequence.}
\end{figure}

The depolarizing noise (or any Pauli errors) on the result qubit anywhere during the protocol can be treated as either the initialization error that reduces $\alpha$ or the measurement error that increases $p$. Errors on the data qubits before the realization of the hidden function does not have any effect since all data qubits are completely mixed, as long as the noise is unital. Also, errors on the data qubits after the controlled-$\tilde{R}_x^{j}(\theta)$ are irrelevant since only the result qubit is detected. In contrast, for $s_j=1$, a phase-flip ($\sigma_z$) error that occurs on a data qubit between the CNOT and the controlled-$\tilde{R}_x^{j}(\theta)$ can propagate to the result qubit.
Then the propagated error can be treated as an error in the state preparation or in the measurement. Because of the properties $\one\otimes\sigma_z\left(|0\rangle\langle 0|\otimes\one+|1\rangle\langle 1|\otimes \sigma_x\right)=\left(|0\rangle\langle 0|\otimes\one+|1\rangle\langle 1|\otimes \sigma_x\right)\sigma_z\otimes \sigma_z$ and $\sigma_z H=H\sigma_x$, two quantum circuits shown in Fig.~\ref{fig:ErrorProp} are equivalent. This shows that a single phase-flip error ($Z$ in Fig.~\ref{fig:ErrorProp}) that occurs on a data qubit results in two errors, a phase-flip and a bit-flip ($X$ in Fig.~\ref{fig:ErrorProp}) on the input state of the data and the result qubits, respectively. 
Now suppose that the phase-flip error corrupts two data qubits simultaneously at this location. This sends two bit-flip errors to the initial state of the result qubit which cancel each other. Hence, the phase errors that occur simultaneously on an even number of data qubits cancel each other and do not affect the result qubit. For an odd number of phase errors, only one of them affects the result qubit. Therefore, the depolarizing noise occurring between the controlled-$U_s$ and the controlled-$\tilde{R}_x^{j}(\theta)$ independently on all data qubits with the error rate $q$ results in a bit-flip error with the error rate being $\sim q/2$ on the initial state of the result qubit. The initial polarization of the result qubit is multiplied by a factor of $\sim (1-q)$.

Systematic errors in the controlled-$\tilde{R}_x^{j}(\theta)$ also affect the result, but not severely. We already mentioned that the algorithm works for all $\theta\neq a\pi,\;a\in\mathbb{Z}$, although ideally $\theta$ should be $\pi/2$ to minimize the computational overhead. Therefore, the algorithm withstands small amplitude errors. It is also robust to the error in the phase of the rotation. For example, consider the rotation around $\cos(\phi)\hat x+\sin(\phi)\hat x_{\perp}$, where $\hat x_{\perp}$ is some axis orthogonal to $\hat x$. Then the normalized trace is multiplied by a factor of $\cos(\phi)^m$. In principle, the algorithm can distinguish $s_j$ as long as $\cos(\phi)\neq 0$, but the optimal separation is attained when $\cos(\phi)=1$ as chosen in our algorithm.
\section{Quantum Discord and Coherence Consumption}
\label{sec:4}
In the preceding, we showed that the learning is enabled by the nonlocal nature of the measurement embedded in each query. This section further investigates the source of the quantum advantage in our protocol from the resource-theoretic standpoint. According to the results in Refs.~\cite{PhysRevA.95.022330,PhysRevA.72.042316}, the DQC1 circuit cannot generate entanglement at the bipartition split as one result qubit and $n$ data qubits when $\alpha\le 1/2$. Hereinafter we limit our discussion to correlations that are generated at this result-data bipartition. Clearly, entanglement is not the source of the quantum supremacy in our algorithm. However, nonclassical correlation other than entanglement as measured by quantum discord can exist for $\alpha>0$~\cite{PhysRevLett.100.050502}. Quantum discord quantifies the quantumness of correlations based on the entropic measure, and it can be understood as the amount of the disturbance induced to a bipartite quantum system via local measurements~\cite{discord_vedral,PhysRevLett.88.017901}.
We examine quantum discord with respect to the measurement on the result qubit in our DQC1 circuit, speculating that it is closely linked to the origin of the quantum advantage. First, the output state in the DQC1 version of the original LPN algorithm (Fig.~\ref{fig:LPNasDQC1}) has zero discord since $U_s^2=\one$ for all $s$~\cite{PhysRevLett.105.190502}. However, discord is generated when the controlled rotation $R_x^{\bar{j}}(\theta)$ is added. We calculate the amount of discord generated in our modified DQC1 circuit shown in Fig.~\ref{fig:modDQC1} for various hidden functions, and it is observed to be different depending on $s_j$. This feature coincides with the dependence of the trace of the total unitary operator on $s_j$ (see Eq.~(\ref{eq:tr_final})), which plays the central role in our learning algorithm.
\begin{figure}[h]
\centering
\includegraphics[width=0.95\columnwidth]{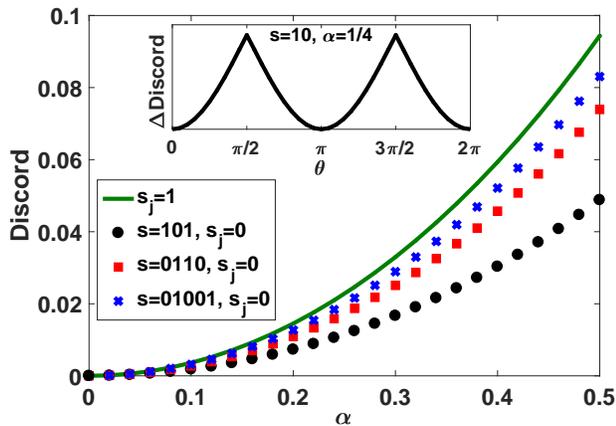}
\caption{\label{fig:DiscordN}Quantum discord at the end of the DQC1 circuit shown in Fig.~\ref{fig:modDQC1} as a function of $\alpha$ for various $s$. In this regime, there is no entanglement in the result-data  bipartition. The amount of discord depends on the hidden bit value encoded in the $j$th data qubit, which is excluded from the uniform rotation prior to the measurement (see Eq.~(\ref{eq:cR})). For all $s$, the discord is the same when $s_j=1$ (solid line). The inset shows the difference of the discord for $s_j=1$ and $s_j=0$ as a function of the controlled-rotation angle $\theta$ when $s=10$ and $\alpha=1/4$.}
\end{figure}
The discord is plotted as a function of $\alpha$ for some selection of the hidden bit strings in Fig.~\ref{fig:DiscordN}. As the length of $s$ increases, the difference of the discords for $s_j=1$ and $s_j=0$ becomes smaller, similar to the behavior of $\Delta\tau_j$. Moreover, the difference of the discords decreases with $\alpha$, consistent with the scaling of the number of queries required in terms of $\alpha$ for a fixed accuracy. The inset shows the difference of the discords for $s_j=1$ and $s_j=0$ as a function of the controlled-rotation angle $\theta$ when $s=10$ and $\alpha=1/4$. The \textit{discord contrast} with respect to $\theta$ resembles $\Delta\tau_j$ in that it is the largest when $\theta$ is an odd-integer multiple of $\pi/2$ and vanishes at the integer multiples of $\pi$ as the discord is zero regardless of $s_j$ at these points.
Above studies suggest that the presence of nonzero discord and the discord contrast in different DQC1 circuits are crucial for our learning algorithm. Nonetheless, claiming quantum discord as the necessary resource for the DQC1-based binary classification in general is problematic since one can come up with two unitary matrices with different normalized traces that do not produce discord when implemented in the DQC1 circuit.

Alternatively, quantum coherence can be regarded as a resource, and it has been rigorously studied within the framework of quantum resource theory recently~\cite{PhysRevLett.113.140401,PhysRevLett.116.120404,RevModPhys.89.041003}. Evidently, the probe qubit must contain some amount of coherence as the minimal requirement for the DQC1 protocol~\cite{2058-9565-1-1-01LT01}. The connection between coherence and discord in DQC1 is established in Ref.~\cite{PhysRevLett.116.160407}: the discord produced is upper-bounded by the coherence consumed by the probe qubit. Using the relative entropy of coherence as the quantifier~\cite{PhysRevLett.113.140401}, the coherence consumption $\Delta C$ in each execution of our DQC1 protocol can be expressed as
\begin{equation}
\Delta C=H_2\left(\frac{1-\alpha\lvert\tau_j\rvert}{2}\right)-H_2\left(\frac{1-\alpha}{2}\right),
\end{equation}
where $H_2(\cdot)$ is the binary Shannon entropy and $\tau_j$ is the normalized trace of the total unitary operator acting on the data qubits controlled by the result qubit.
This is monotonically decreasing with respect to $\lvert\tau_j\rvert$ for a fixed nonzero $\alpha$. Thus it appears that the DQC1 protocol is inherently capable of quantifying the consumption of the coherent resource supplied by one quantum bit. Furthermore, the magnitude of the partial derivative of $\Delta C$ with respect to $\lvert\tau_j\rvert$ ($\alpha$) monotonically increases with the independent variable, meaning that $\Delta C$ is more sensitive to the changes in $\lvert\tau_j\rvert$ ($\alpha$) when the independent variable is large.
This feature is consistent with the computational complexity of our algorithm. By all means, the notion of the coherence consumption is purely quantum mechanical. Our algorithm is set up in a way that the coherent resource used up in each query varies with the answer $s_j$ being probed. An interesting open question is whether manipulating the coherence consumption provides a quantum advantage in solving problems other than those based on the trace estimation.
\section{Conclusion}
\label{sec:5}
By measuring only one quantum bit with nonzero polarization in each query, an $n$-bit hidden parity function can be identified. This situation arises when data qubits undergo the completely depolarizing channel in the original quantum LPN algorithm in Ref.~\cite{LPNTheory}. The protocol introduced here can solve the problem efficiently when $n\sim \log_2(L)$. Classically, the corresponding task can only be accomplished via brute-force enumeration in an exponentially large search space, provided that an efficient means to verify the answer exists. The one-qubit LPN algorithm is inspired by the DQC1 model. However, the naive translation of the original LPN algorithm to a DQC1 circuit does not solve the problem since the trace of the unitary matrix that encodes the hidden parity function is zero in $2^n-1$ instances. To circumvent the issue, we introduced controlled uniform rotations so that the trace is either zero or nonzero depending on the hidden bit value encoded in the data qubit being probed. The additional operation can be viewed as the nonlocal measurement between the result and the data qubit.
The mere existence of nonzero quantum discord between the result and data qubits does not permit the learning. Instead, we conjectured that the discord contrast or, more fundamentally, the \textit{coherence consumption contrast} is essential for the quantum advantage in our algorithm.

While efforts towards building standard quantum computers that fulfill what the theory of QIP promises continue, exploring weaker but more realistic quantum devices to solve interesting but classically hard problems is imperative. The LPN problem is one such problem in which the noisy quantum machine can shine. For the LPN problem, the ability to manipulate and measure the coherence consumed by one quantum bit suffices to demonstrate the quantum supremacy. This also motivates future studies on whether similar strategies can be utilized in the near-term quantum devices to perform other well-defined computational tasks beyond classical capabilities and how much, if any, improvement can be achieved by utilizing coherence from more than one qubit.
\begin{acknowledgments}
We thank Sumin Lim for helpful discussions. This research was supported by the National Research Foundation of Korea (Grants No. 2015R1A2A2A01006251 and No. 2016R1A5A1008184).
\end{acknowledgments}


\begin{thebibliography}{10}

\bibitem{Nigg302}
D.~Nigg, M.~M{\"u}ller, E.~A. Martinez, P.~Schindler, M.~Hennrich, T.~Monz,
  M.~A. Martin-Delgado, and R.~Blatt, 
  Science \textbf{345}, 302 (2014).

\bibitem{NVQEC}
T.~H. Taminiau, J.~Cramer, T.~van~der Sar, V.~V. Dobrovitski, and R.~Hanson, 
Nat. Nanotechnol. \textbf{9}, 171 (2014).

\bibitem{SCQEC}
J.~Kelly, R.~Barends, A.~G. Fowler, A.~Megrant, E.~Jeffrey, T.~C. White,
  D.~Sank, J.~Y. Mutus, B.~Campbell, Y.~Chen, Z.~Chen, B.~Chiaro, A.~Dunsworth,
  I.~C. Hoi, C.~Neill, P.~J.~J. O'Malley, C.~Quintana, P.~Roushan,
  A.~Vainsencher, J.~Wenner, A.~N. Cleland, and J.~M. Martinis, 
  Nature (London) \textbf{519}, 66 (2015).

\bibitem{NMR2017}
D.~Lu, K.~Li, J.~Li, H.~Katiyar, A.~J. Park, G.~Feng, T.~Xin, H.~Li, G.~Long,
  A.~Brodutch, J.~Baugh, B.~Zeng, and R.~Laflamme, 
  npj Quantum Info. \textbf{3}, 45 (2017).

\bibitem{Angluin1988}
D.~Angluin and P.~Laird, Mach. Learn. \textbf{2}, 343 (1988).

\bibitem{Blum2003}
A.~Blum, A.~Kalai, and H.~Wasserman, J. Assoc. Comput. Mach. \textbf{50}, 506 (2003).

\bibitem{Lyubashevsky2005}
V.~Lyubashevsky, ``The Parity Problem in the Presence of Noise, Decoding
  Random Linear Codes, and the Subset Sum Problem," 
  in {\em Approximation, Randomization and Combinatorial Optimization: Algorithms and Techniques}, Lecture Notes in Computer Science, 
  Vol.~3624 (Springer, Berlin, 2005), pp.~378--389.

\bibitem{Levieil2006}
{\'E}.~Levieil and P.-A. Fouque, ``An Improved LPN Algorithm",
 in {\em Security and Cryptography for Networks: SCN 2006}, Lecture Notes in Computer Science, 
 Vol.~4116 (Springer, Berlin, 2006), pp.~348--359.

\bibitem{Regev:2005}
O.~Regev, in {\em Proceedings of the Thirty-Seventh Annual ACM
  Symposium on Theory of Computing, STOC'05} (ACM, New York, 2005), pp.~84--93.

\bibitem{Pietrzak2012}
K.~Pietrzak, ``Cryptography from Learning Parity with Noise,"
 in {\em Theory and Practice of Computer Science, SOFSEM 2012}, 
 Vol.~7147 (Springer, Berlin, 2012), pp.~99--114.

\bibitem{LPNTheory}
A.~W. Cross, G.~Smith, and J.~A. Smolin,
 Phys. Rev. A \textbf{92}, 012327 (2015).

\bibitem{LPNexp}
D.~Rist{\`e}, M.~P. da~Silva, C.~A. Ryan, A.~W. Cross, A.~D. C{\'o}rcoles,
  J.~A. Smolin, J.~M. Gambetta, J.~M. Chow, and B.~R. Johnson,
  npj Quantum Info. \textbf{3}, 16 (2017).

\bibitem{DQC1PhysRevLett.81.5672}
E.~Knill and R.~Laflamme,
 Phys. Rev. Lett. \textbf{81}, 5672 (1998).

\bibitem{PhysRevA.72.042316}
A.~Datta, S.~T. Flammia, and C.~M. Caves,
 Phys. Rev. A \textbf{72}, 042316 (2005).

\bibitem{DQC1complexity}
P.~W. Shor and S.~P. Jordan,
 Quantum Info. Comput. \textbf{8}, 681 (2008).

\bibitem{estimate}
P.~J. Huber, {\em Robust Statistics} (Wiley, New York, 1981).

\bibitem{PhysRevA.95.022330}
M.~Boyer, A.~Brodutch, and T.~Mor,
 Phys. Rev. A \textbf{95}, 022330 (2017).

\bibitem{PhysRevLett.100.050502}
A.~Datta, A.~Shaji, and C.~M. Caves,
 Phys. Rev. Lett. \textbf{100}, 050502 (2008).

\bibitem{discord_vedral}
L.~Henderson and V.~Vedral,
 J. Phys. A: Math. Gen. \textbf{34}, 6899 (2001).

\bibitem{PhysRevLett.88.017901}
H.~Ollivier and W.~H. Zurek,
 Phys. Rev. Lett. \textbf{88}, 017901 (2001).

\bibitem{PhysRevLett.105.190502}
B.~Daki\'{c}, V.~Vedral, and \v{C}. Brukner,
 Phys. Rev. Lett. \textbf{105}, 190502 (2010).

\bibitem{PhysRevLett.113.140401}
T.~Baumgratz, M.~Cramer, and M.~B. Plenio,
 Phys. Rev. Lett. \textbf{113}, 140401 (2014).

\bibitem{PhysRevLett.116.120404}
A.~Winter and D.~Yang,
 Phys. Rev. Lett. \textbf{116}, 120404 (2016).

\bibitem{RevModPhys.89.041003}
A.~Streltsov, G.~Adesso, and M.~B. Plenio,
 Rev. Mod. Phys. \textbf{89}, 041003 (2017).

\bibitem{2058-9565-1-1-01LT01}
J.~M. Matera, D.~Egloff, N.~Killoran, and M.~B. Plenio, 
 Quantum Science and Technology \textbf{1}, 01LT01 (2016).

\bibitem{PhysRevLett.116.160407}
J.~Ma, B.~Yadin, D.~Girolami, V.~Vedral, and M.~Gu,
 Phys. Rev. Lett. \textbf{116} 160407 (2016).

\end{thebibliography}
\end{document}